\documentclass[conference]{IEEEtran}
\IEEEoverridecommandlockouts
\usepackage{todonotes}

\usepackage[style=numeric,citestyle=numeric-comp,backend=biber,sorting=none,maxbibnames=99]{biblatex}
\usepackage{amsmath,amssymb,amsfonts}
\usepackage{algorithmic}
\usepackage{graphicx}
\usepackage{textcomp}
\usepackage{xcolor}
\usepackage{caption}
\usepackage{subcaption}
\usepackage{printlen}
\begin{document}

\newcommand{\filledcircle}{%
  \tikz{
    \fill[gray] (0,0) circle (0.6ex);
    \useasboundingbox (-0.6ex,-0.6ex) rectangle (0.6ex,0.6ex);
  }%
}

\newcommand{\markercross}[1][black]{%
  \tikz{
    \draw[line width=1.4pt, gray]
      (-0.6ex,-0.6ex) -- (0.6ex,0.6ex)
      (-0.6ex,0.6ex) -- (0.6ex,-0.6ex);
    \useasboundingbox (-0.6ex,-0.6ex) rectangle (0.6ex,0.6ex);
  }%
}

\title{Secure and Scalable Rerouting in LEO Satellite Networks}

\author{
  \IEEEauthorblockN{Lyubomir Yanev}
  \IEEEauthorblockA{ETH Zürich\\ lyanev@ethz.ch}
  \and
  \IEEEauthorblockN{Pietro Ronchetti}
  \IEEEauthorblockA{ETH Zürich\\ pietroro@ethz.ch}
  \and
  \IEEEauthorblockN{Joshua Smailes}
  \IEEEauthorblockA{University of Oxford\\ joshua.smailes@cs.ox.ac.uk}

  \and
  \IEEEauthorblockN{Martin Strohmeier}
  \IEEEauthorblockA{armasuisse Science + Technology\\ martin.strohmeier@armasuisse.ch}

}

\maketitle

\begin{abstract}
Resilient routing in large-scale Low Earth Orbit (LEO) satellite networks remains a key challenge due to frequent and unpredictable link and node failures, potentially in response to cybersecurity breaches. While prior work has explored rerouting strategies with various levels of network awareness, their relative tradeoffs under dynamic failure conditions remain underexplored. 

In this work, we extend the Deep Space Network Simulator (DSNS) to systematically compare three rerouting paradigms, each differing in the scope of failure knowledge available to each node. We compare local \textit{neighbor-based}, \textit{segment-based} and \textit{global-knowledge-based} rerouting as well as a naive \textit{source routing} solution that is unaware of failures.

Our main goal is to evaluate how the breadth of failure awareness impacts routing performance and resilience under failures, both random and targeted. We measure delivery ratio, latency, rerouting overhead, and loop occurrence. Our findings show the potential of \textit{segment-based rerouting} to achieve a favorable tradeoff between local responsiveness and global coordination, offering resilience benefits with minimal overhead—insights that can inform future fault-tolerant satellite network design.

\end{abstract}

\section{Introduction}

The rapid expansion of Low Earth Orbit (LEO) satellite constellations—such as Starlink, OneWeb, and Kuiper—has revitalized interest in large-scale, space-based networking. These networks offer global coverage and low-latency communication, but also introduce challenges such as frequent handovers and limited onboard resources. Importantly, they are susceptible to link/node failures that can be both random and targeted by an adversary using anti-satellite weapons \cite{zhang2023fast,willbold2023space,pavur2019cyber}. In such dynamic environments, even small disruptions can break precomputed paths and cause significant degradation in delivery performance.

Existing works on satellite routing emphasize static topologies, periodic updates, and fast reroute (FRR) mechanisms tuned for single-link failures (e.g., IPFRR, Loop-Free Alternates \cite{shand2010ip, atlas2008basic}). These approaches rely on stable grid-like structures, geographic partitions, and precomputed local backup paths.

However, maintaining precomputed backup paths becomes inefficient or impractical when failures are unpredictable and widespread. Such conditions are increasingly likely in large, dynamic constellations where the ability to deal with multi-link failures becomes increasingly important \cite{zhang2023fast}. One possible approach, disseminating control signaling globally through the entire network, causes significant overhead and data loss \cite{jin2022disruption}. Therefore, limiting the signaling overhead is also necessary.

Thus, we present an approach tailored to the real-world constraints and challenges of satellite networks, where frequent topology changes and limited onboard memory make persistent global state impractical and quickly render precomputed paths obsolete. In this paper, we propose a fresh take on \textit{segment-based rerouting} in satellite networks, which avoids reliance on the underlying network topology when making routing decisions and provides improved failure awareness. We compare it with a localized \textit{neighbor-based rerouting} approach, pure \textit{source routing} and a globally optimal failure awareness baseline.

We provide a comprehensive evaluation of the effects that the scope of failure knowledge available to satellites has on routing outcomes. To isolate this effect, we use a uniform routing design: no routing tables, no precomputed paths, and no global state dissemination. All decisions are made on-demand, using available knowledge of failures at the time of forwarding.
We extend the Deep Space Network Simulator (DSNS)~\cite{smailes2025dsns}, originally developed for public-key infrastructure (PKI) message transport in delay-tolerant space environments~\cite{smailesKeySpace2025}, by implementing the rerouting paradigms described above. We simulate all four under baseline and fault conditions, including randomized multi-link disruptions and targeted attacks on structurally-important nodes.

Our findings demonstrate a clear trend: as failure-awareness increases, latency, path quality, and resilience metrics improve progressively. However, this improvement comes with corresponding increases in signaling overhead, computation and memory, highlighting the need to balance failure awareness scope with operational cost.\\

Our contributions in this paper are as follows:

\begin{itemize}
    \item We extend the Deep Space Network Simulator (DSNS) to implement four routing paradigms — pure \textit{source routing}, \textit{neighbor-based rerouting}, \textit{segment-based rerouting}, and \textit{global-knowledge-based rerouting}.

    \item We conduct a systematic simulation-based evaluation across three large-scale constellations (Iridium, Starlink, and LEO/LEO), injecting both random and targeted failures. We find that segment-based rerouting achieves up to 30\% fewer message drops and significantly reduces routing loops compared to neighbor-based rerouting under high failure rates.

    \item We identify key security-relevant tradeoffs between failure-awareness scope, delivery performance, and signaling overhead. \textit{Segment-based rerouting} provides a scalable and attack-resilient compromise, with lower latency than source routing and up to 80\% fewer loops per message than neighbor-based approaches.
\end{itemize}

\section{Related Work}

Recent research on LEO satellite networks spans a wide range of topics, including routing design, fault tolerance and simulation environments. In this section, we present an overview of relevant work spanning four main themes: stateless and on-demand routing, satellite network partitioning and zone-based approaches, routing resilience and failure modeling, and simulation frameworks for performing experiments on satellite networks.

\subsection{Stateless and On-Demand Routing} 
The authors in \cite{vissicchio2024reliable} propose StarGlider, a stateless forwarding model in which satellites forward based on minimal state and do not compute paths. It delegates all routing decisions to ground stations and thus avoids keeping any routing state or performing computation in orbit.

We share similar constraints---our satellites keep no routing tables and do not exchange routing paths. However, unlike StarGlider, we allow lightweight, on-demand, scoped path computation when failures are encountered. This hybrid model supports responsiveness to unpredictable failures while preserving scalability. 

Precomputed backup paths have been used by fast reroute schemes such as IPFRR \cite{shand2010ip}, and, more recently, by \cite{chen2023fast, du2025fast}. Although they successfully solve the problem of single-link failures, they generally become inefficient when presented with the more complex problem of multi-link failures, another intricacy of LEO satellite networks \cite{zhang2023fast}.

Our on-demand routing model directly addresses multi-link failures when making decisions, flexibly adapting paths using current failure awareness.

\subsection{Network Partitioning and Zone-Based Approaches}
Segmented or zone-based designs have been explored in an attempt to alleviate the burden of signaling/communication overhead from satellite networks.

Schemes like ZDRP \cite{dong2011zone} and ASER \cite{zhang2021aser}  aim to avoid unnecessary flooding and increase recovery speed by confining control signaling to bounded domains. These typically rely on precomputed backup paths or geographic/topological zones making use of the specific topology of the network. ASER, for example, partitions the network using grid-based areas and maintains routing tables within each. In contrast to those approaches, our model does not explicitly rely on the underlying network topology, neither for partitioning the network nor for making routing decisions.

\subsection{Routing Resilience and Failure Model}
Several works analyze robustness in satellite networks by simulating failures and studying their impact on delivery or connectivity. The authors in \cite{li2024analyzing,zhang2021aser} model random and load-based failures to evaluate robustness under attack, while \cite{yang2020enhancing} and \cite{feng2020elastic} explore recovery mechanisms and the importance of elastic response strategies. These motivate our own use of random and targeted failure models, particularly the high-centrality node failures used to test segment-based rerouting under stress. 

To our knowledge, however, there exists no comprehensive analysis of the effect that the scope of failure awareness has on performance, resilience, and recovery from failures in different routing approaches and satellite network types.

\subsection{Simulation Frameworks}

A number of simulation frameworks have been proposed to support research on LEO satellite networks. \cite{manzanares2025comprehensive} provides a detailed overview of two such tools based on the ns-3 simulator---Hypatia and ns-3-leo.
Whilst Hypatia \cite{kassing2020exploring} makes use of static routing and does not explicitly model failure-aware behavior, ns-3-leo employs a more dynamic model, but still keeps its focus on performance metrics and routing protocols. StarryNet \cite{lai2023starrynet} enables real-data-driven simulations of LEO constellations, building a virtual representation of a physical satellite network. We also evaluate the impact of failures on performance, but our focus is elsewhere: we investigate the impact that the awareness of the failures has on failure mitigation strategies.

To this end, we use DSNS \cite{smailes2025dsns} as the base simulation framework. We extend DSNS with failure injection and scoped rerouting capabilities tailored specifically to evaluating robustness of satellite networks under dynamic failures.

\section{System Model}

\begin{figure*}[t]
    \centering
    \begin{subfigure}[t]{0.40\textwidth}
        \centering
        \includegraphics[width=\linewidth]{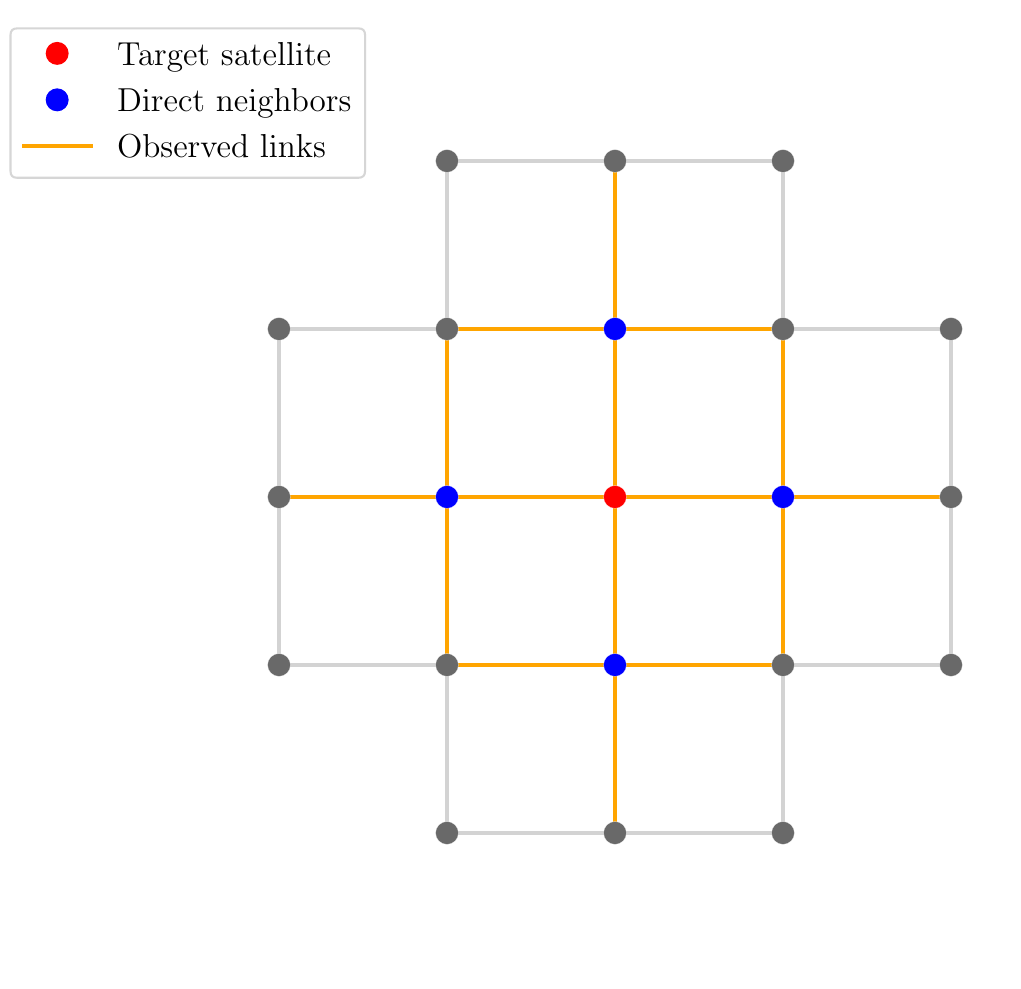}
        \caption{Satellites are aware not only of the status of their adjacent links, but also those adjacent to their neighbors.}
        \label{fig:awareness-neighbor}
    \end{subfigure}
    \hfill
    \begin{subfigure}[t]{0.55\textwidth}
        \centering
        \includegraphics[width=\linewidth]{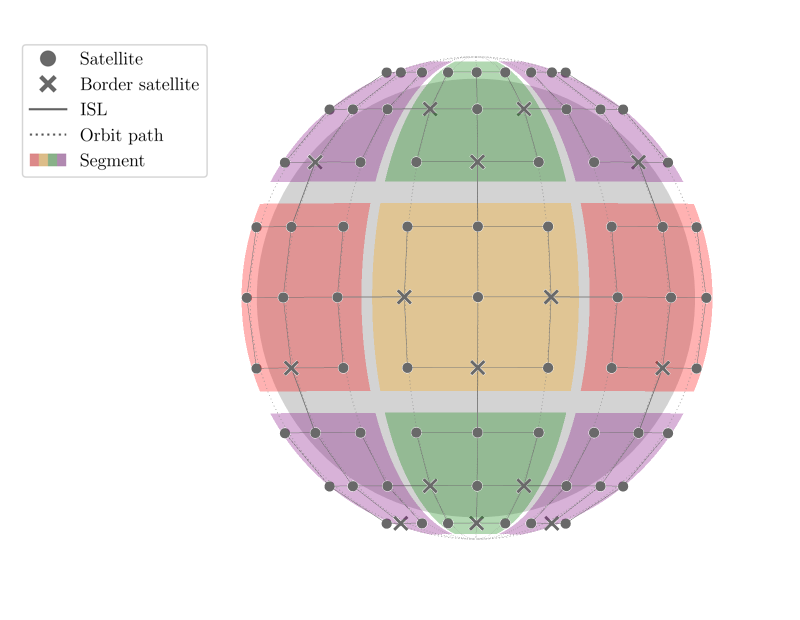}
        \caption{Satellites are aware of all link failures within their respective segment. This information is relayed from border satellites (\markercross) to all satellites within their segment (\filledcircle)}
        \label{fig:awareness-segment}
    \end{subfigure}
    \caption{Failure awareness in neighbor- and segment-based rerouting.}
    \label{fig:failure-awareness-scopes}
\end{figure*}

In this section, we first outline the core modeling assumptions inherited from the DSNS simulator and those specific to our experimental setup. Then, we describe our traffic generation and failure models.

\subsection{Modeling Assumptions}

We build on DSNS, which utilizes store-and-forward lookahead routing, and for our purposes serves as a baseline for event-based Delay Tolerant Networking. Concretely, we make five main assumptions:

\begin{enumerate}
    \item We assume that only predictable topology knowledge (such as future contact windows) is taken into consideration using a lookahead approach, as in the original DSNS implementation. We therefore also assume that ground stations can precompute low-delay paths based on scheduled contact windows. Failure information is acquired in real-time according to the failure-awareness strategy, not predicted in advance.

    \item We assume multi-orbital satellite constellations with multiple polar orbital planes and inter-satellite links (ISLs) defined based on visibility and elevation angle threshold. Ground stations connect opportunistically based on geographic position and satellite visibility.

    \item We assume on-demand routing, where nodes recompute paths within their knowledge scope (either local neighborhood, segment, or global) upon detecting a failure. If no failure is detected in the message's path, the satellite simply sends the message onwards according to its precomputed path.

    \item We assume that the paths computed when satellites have to reroute are optimal, as far as their failure knowledge allows. That is, a node with global knowledge uses the latest known global topology to compute the best path, whereas a node with only neighboring or segment knowledge operates with limited topological visibility, either in a local radius or within a predefined segment respectively (cf. Fig.~\ref{fig:failure-awareness-scopes}). This hybrid design allows responsiveness to real-time disruptions while avoiding unnecessary flooding of failure information.

    \item Our routing decisions generally do not assume geometric regularity or orbital symmetry. One of the aims of this approach is that as long as the network remains connected, it is possible to recover from multi-link random failures.

\end{enumerate}

\subsection{Traffic Model}

Our traffic model simulates simple message flows between randomly selected ground station pairs. These messages could represent typical space network control-plane tasks such as certificate synchronization, key updates, or routing metadata propagation. Traffic is generated in very frequent bursts, while failures are uniform, random and continuous throughout the whole duration. Message deliveries are independent of each other and thus this model preserves statistical validity. Messages are routed exclusively through the ISL (space) network.

\subsection{Failure Model and Attack Scenarios}
In our simulations, we consider two classes of failures, random and targeted:

\begin{enumerate}

\item \textbf{Random Failures:}
    We simulate dynamic and unpredictable disruption patterns, injecting failures throughout the duration of the simulation by selectively and continuously removing links for a specified downtime, via events. On average, a set fraction of the network's links is kept down at all times. 
    
\item \textbf{Targeted Failures:} 
    Select nodes or links are disabled either permanently or for a specified downtime. In our case, we target segment-based rerouting by selectively disabling border satellites.

\end{enumerate}

We assume that border satellites are immune to random failures and are only affected by targeted attacks. This simplification helps isolate the behavior of inter-segment routing and border satellite effectiveness and analyze more concretely how a widened failure scope helps performance and efficiency.

\section{Space Routing Paradigms}
\begin{figure*}[t]
    \centering
    \begin{subfigure}[b]{0.46\textwidth}
        \centering
        
        \includegraphics[width=\linewidth]{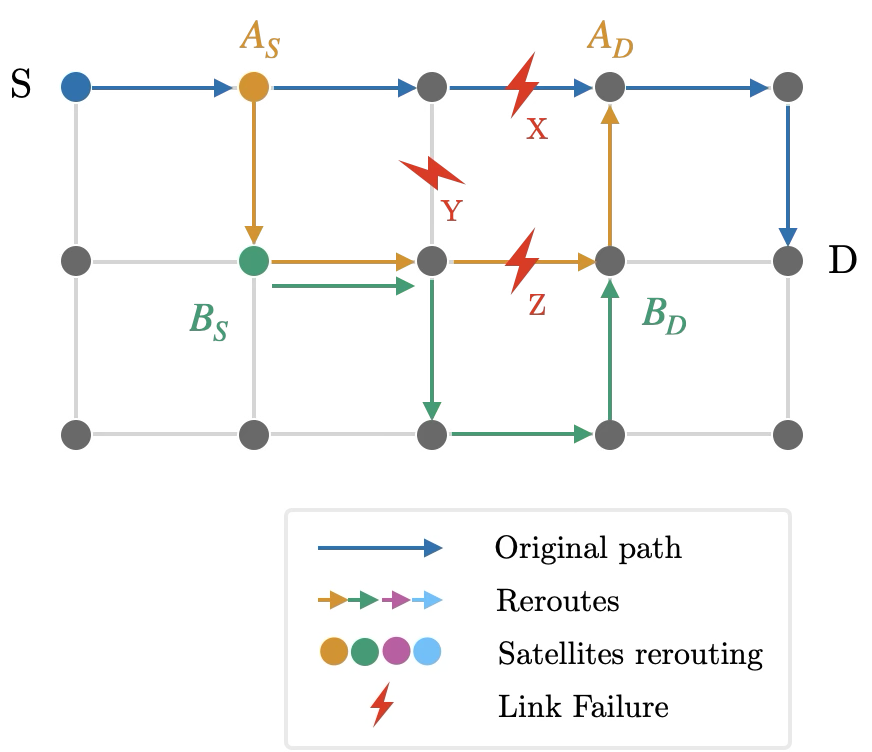}
        \vspace*{0mm}
        \caption{The original path goes through failure X. While S is unaware of this failure, $A_S$ is. Using its knowledge of X and Y, $A_S$ is able to reroute around the failure towards $A_D$. Similarly, $B_S$ is aware of Z, a failure on the rerouted path. It reroutes the modified path towards $B_D$.}
        \label{fig:sub-a}
    \end{subfigure}
    \hfill
    \begin{subfigure}[b]{0.46\textwidth}
        \centering
        
        \includegraphics[width=\linewidth]{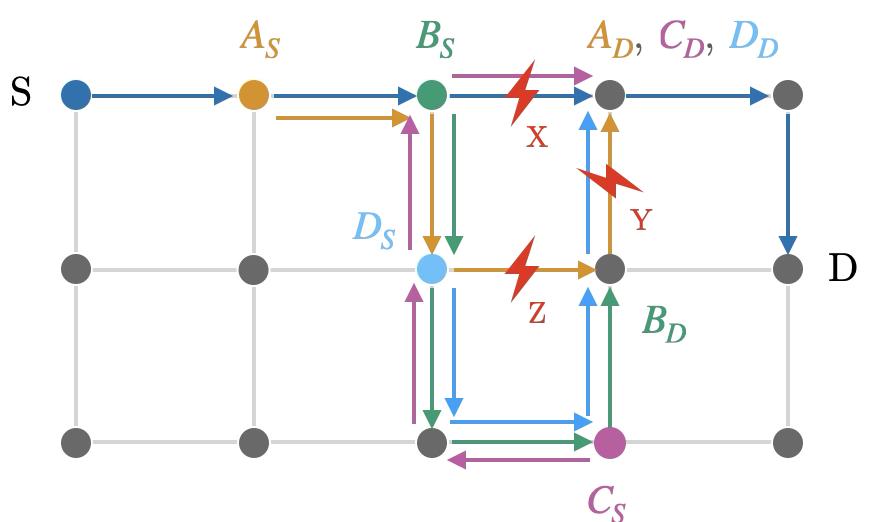}
       
        \vspace*{8.5mm}
        \caption{Similarly to subfigure~\subref{fig:sub-a}, $A_S$ notices the original path is blocked by failure X and uses this information to reroute towards $A_D$, unknowingly through failure Z. $B_S$ is aware of Z, and thus reroutes the message to $B_D$. Crucially, $B_S$ is unaware of failure Y due to X cutting off communication with its neighbor. $C_S$ is the first node to be aware of Y and tries to reroute towards $C_D$, but does not know about X. $D_S$, while being aware of X, is unaware of Y, and a routing loop is formed when it tries to reroute $C_S$'s route to $D_D$.}
        \label{fig:sub-b}
    \end{subfigure}
    \caption{Handling failures in the neighbor-based rerouting paradigm}
    \label{fig:combined}
\end{figure*}

We implement and compare four routing paradigms, each characterized by a different failure-awareness scope and rerouting capability. These paradigms differ in how much knowledge each node has about the state of the network, how rerouting decisions are made, and how resource intensive they are. Crucially, all paradigms operate on-demand: a node only acquires failure-related information when attempting to forward a message, and does not continuously update or propagate global state. Failed links that a satellite is aware of are not considered when rerouting since they are excluded from its view of the graph. To prevent infinite message loops under rerouting, each message tracks its visited nodes and is dropped if it exceeds a constellation-specific loop threshold. If rerouting fails entirely, the message is stored at the first failed link that was present on the message's path prior to the reroute attempt.

\subsection{Pure Source Routing} 
Pure source routing is the baseline approach: a fixed path from sender Ground Station Terminal (GST) to receiver GST is computed at message creation time using a snapshot of the current topology. Once selected, the path is embedded in the message, and no rerouting occurs---even if links fail along the way. This approach incurs no runtime computation overhead but performs poorly under failures, especially if links fail for an extended period of time, or permanently.

\subsection{Neighbor-Based Rerouting} 

When a link failure occurs in the neighbor-based rerouting paradigm, both nodes on the link immediately alert all neighbors to which they have an active link. We assume no delays in this alert due to the immediate closeness of the neighbors. Thus, each satellite is aware of the status of all links up to two hops away (see Fig.~\ref{fig:awareness-neighbor}). This approach is inspired in part by fast reroute schemes such as \cite{chen2023fast} and a very similar strategy employed in \cite{jin2022disruption}, limiting link failure propagation. Failure awareness is only shared between each neighboring satellite and therefore uses little memory.

\paragraph*{Handling Link Failures}

If the receiving satellite determines that the message is currently set to travel along a path that includes a link failure (of which the node is aware), the node attempts to find an alternative path to the other side of the failed link. It then combines this new path with the remaining portion of the original path, utilizing its failure knowledge and local topology information (see Fig.~\ref{fig:sub-a}) — a very constrained view of the graph, also having a small memory footprint. Messages are allowed to backtrack when being rerouted. 

This paradigm represents a minimal-awareness, reactive approach that avoids global updates but may at times lead to suboptimal paths (see Fig.~\ref{fig:segment-vs-neighbor}) or local loops (see Fig.~\ref{fig:sub-b}).

\subsection{Segment-Based Rerouting} 

In segment-based rerouting, the network is first statically partitioned into a configurable number of segments, using a k-medoids clustering approach inspired by the improved k-means algorithm employed in \cite{dai2021intelligent} (see Fig. \ref{fig:awareness-segment}). It uses a distance matrix taking pairwise end-to-end latency between nodes into account, based on a snapshot of the network topology at the start of the simulation. These segments are virtual and fixed throughout the simulation.

Our static partitioning assumption is justified by the stability of LEO satellite neighborhoods: although satellite constellations are inherently mobile, Walker-based constellations retain their neighborhoods very well (i.e., local neighborhoods of satellites remain stable over time). This makes static segmentation a viable and low-overhead choice in our simulation framework.

Satellites are exclusively forced to route only within the segment that they are in.\\

\paragraph{Border Satellites Selection}

We define border satellite candidates as nodes that have links to multiple segments. Fig.~\ref{fig:awareness-segment} shows a simple allocation of border satellites in the network. One is chosen per pair of adjacent segments, based on its closeness centrality relative to the nodes within both segments it connects.

Note that inter-segment links are also relatively stable across time, making the initial selection suitable for the whole length of the simulation under the static partitions assumption. 
These border satellites act as the hubs for failure knowledge---each border satellite stores link failure information about both segments it connects. The border satellites thus maintain knowledge about a larger slice of the network, requiring more memory than the neighbors paradigm. \\

\paragraph{Initial Path Calculation} When sending a message, the GST uses Dijkstra's algorithm to calculate the path of the message to its destination (with failure knowledge of only the segment it is closest to), and then converts the path into a sequence of segments that are to be traversed along the message's path. Then, if the message's destination is within the same segment as its source, it simply computes the optimal path to the destination. Otherwise, using the sequence of segments on the message's path, it determines which border satellites to send the message through, in sequence, such that the final border satellite it is sent to forwards it to its destination.\\

\paragraph{Handling Link Failures}

\begin{figure}
    \centering
    \includegraphics[width=0.4\textwidth]{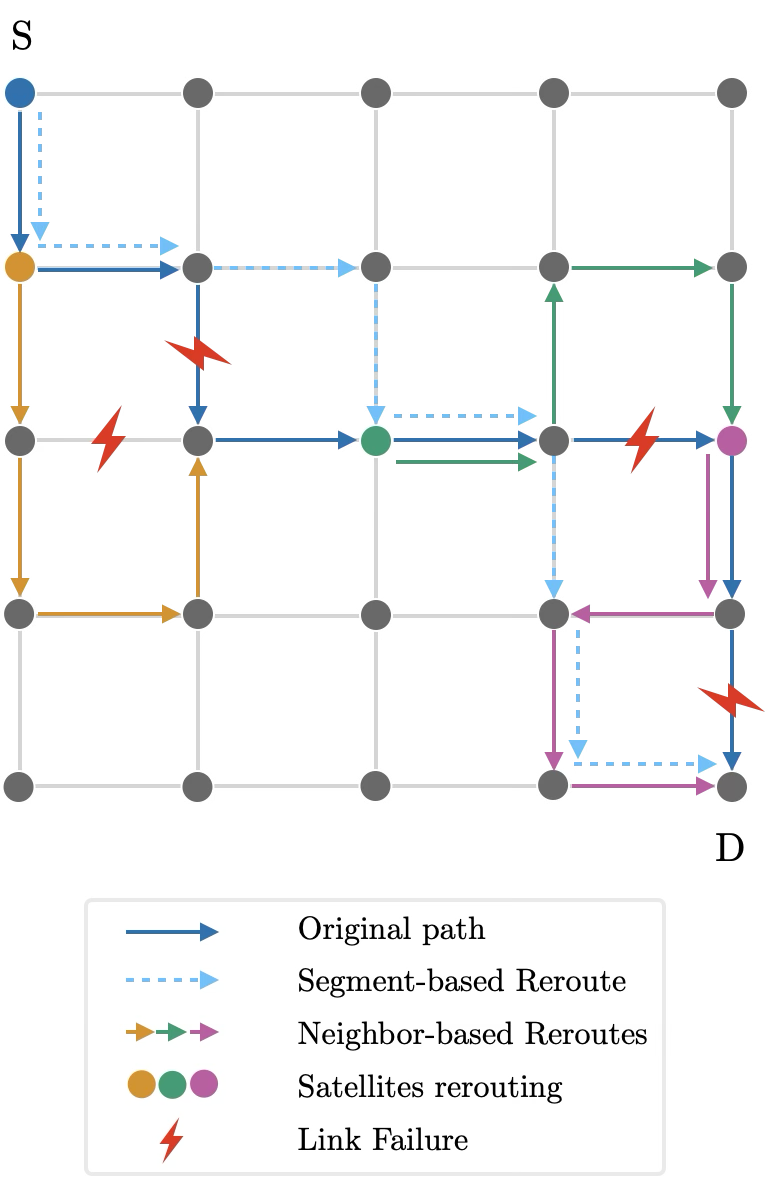}
    \vspace*{2mm}
    \caption{In segment-based rerouting, messages experience fewer reroute attempts due to wider failure awareness within the segment. Failures shown are those present at time of message path calculation.}
    \label{fig:segment-vs-neighbor}
\end{figure}
When a link failure occurs, the information of this failure is propagated to all border satellites of the segment where the failure occurred.

Upon receiving a message, the satellite accesses the failure knowledge from its closest border satellite (which comes with a delay), and then determines whether a reroute is necessary. If it is, the message is rerouted to its next destination: its next border satellite (if its final destination is not within the receiving satellite's segment), or directly to its destination. In both cases this uses the full knowledge of the segment subnetwork. 
In our current implementation, the signaling delay is only added to a message's latency when the knowledge gained was actually used to perform a reroute, so that we remain consistent with our assumption that the paths chosen in each failure awareness level are maximally optimal as far as their failure awareness allowed.

Each node is thus aware of the topology within its own segment on-demand. Furthermore, as knowledge of link failures propagates only to the relevant border satellites, and only when queried during message forwarding, no unnecessary flooding occurs. Fig.~\ref{fig:segment-vs-neighbor} illustrates this: in segment-based rerouting, the GST leverages broader failure awareness when computing the path, avoiding failed links from the start. By contrast, the neighbor-based approach lacks visibility of distant failures, producing a less optimal path that triggers more rerouting. Consequently, messages run a higher risk of loops in neighbor-based rerouting.

\subsection{Global Rerouting} 
Global rerouting is intended to be as close to an optimal case as possible, providing an upper bound to compare against. All nodes possess immediate awareness of any link failure within the whole network. Therefore, the paths chosen here are optimal under complete failure awareness. Nodes use this awareness when determining whether a reroute is necessary, and upon detecting a failed link in a message's path, a receiving node can immediately reroute using the most up-to-date information about failed links throughout the whole network. While this enables optimal paths and fast recovery, it imposes a significant (and unrealistic) communication and computation overhead. 

\subsection{The Routing Cost of Incorrect Failure Assumptions}
One key concern of on-demand routing is its sensitivity to \textit{false positives} in failure detection. If a node mistakenly believes a link is down---e.g., by querying failure knowledge just before the link recovers—it may avoid a viable path, leading to inflated latency, unnecessary reroutes, or dropped messages. Since routing decisions are made on-demand, such stale assumptions persist until explicitly corrected.

\section{Experiment Design}

We conduct simulations using three constellations: an Iridium constellation with 66 satellites, a Starlink constellation with 1584 satellites, and a LEO/LEO constellation with 2 layers and 1650 satellites in total. 

In segment-based rerouting, each constellation is partitioned into 3 segments. Each configuration has 256 ground stations, and their placement remains fixed across simulations. In both failure scenarios explained in the following, messages are independent and generated in bursts, with uniform random emission over short intervals (3 seconds per burst), approximating an average sustained rate of 1 message per second for Iridium, and 1 message per 2 seconds for Starlink and LEO/LEO.

\subsection{Random Failures} 
In the random failure model, failures are injected continuously throughout the simulation such that a fixed fraction (0\%, 15\%, or 30\%) of links are kept failed at all times, each link being kept down for 60 seconds before recovering. Thus, we ensure links fail whilst messages are in transit, underlining our focus on dynamic failure recovery. The simulations run for 7200 seconds for Iridium and 1800 seconds for Starlink and LEO/LEO.

\subsection{Targeted Failures} 
In the targeted failure model, we use the segment-based rerouting and disable all border satellites for a specified amount of time. We then observe the effects this has on throughput and message drops.
We run simulations on the Iridium, Starlink and LEO/LEO constellations for a duration of 1200 seconds, with a message being sent every second. The satellite network is partitioned into 3 segments. All border satellites are disabled at 450 seconds for a duration of 300 seconds, after which they become operational again.

\section{Results}

We run experiments across all rerouting paradigms and constellations, analyzing delivery and drop rates, latency and throughput under both normal and failure conditions. Under random failures, we also explore the prevalence and severity of routing loops, including how often messages revisit previously visited nodes. 

In doing this, we quantify the tradeoffs between resilience, responsiveness, and overhead as failure-awareness scope increases and link failure rate varies.

\subsection{Random Failures} 

\begin{figure}[t]
    \centering
    \includegraphics[width=0.5\textwidth]{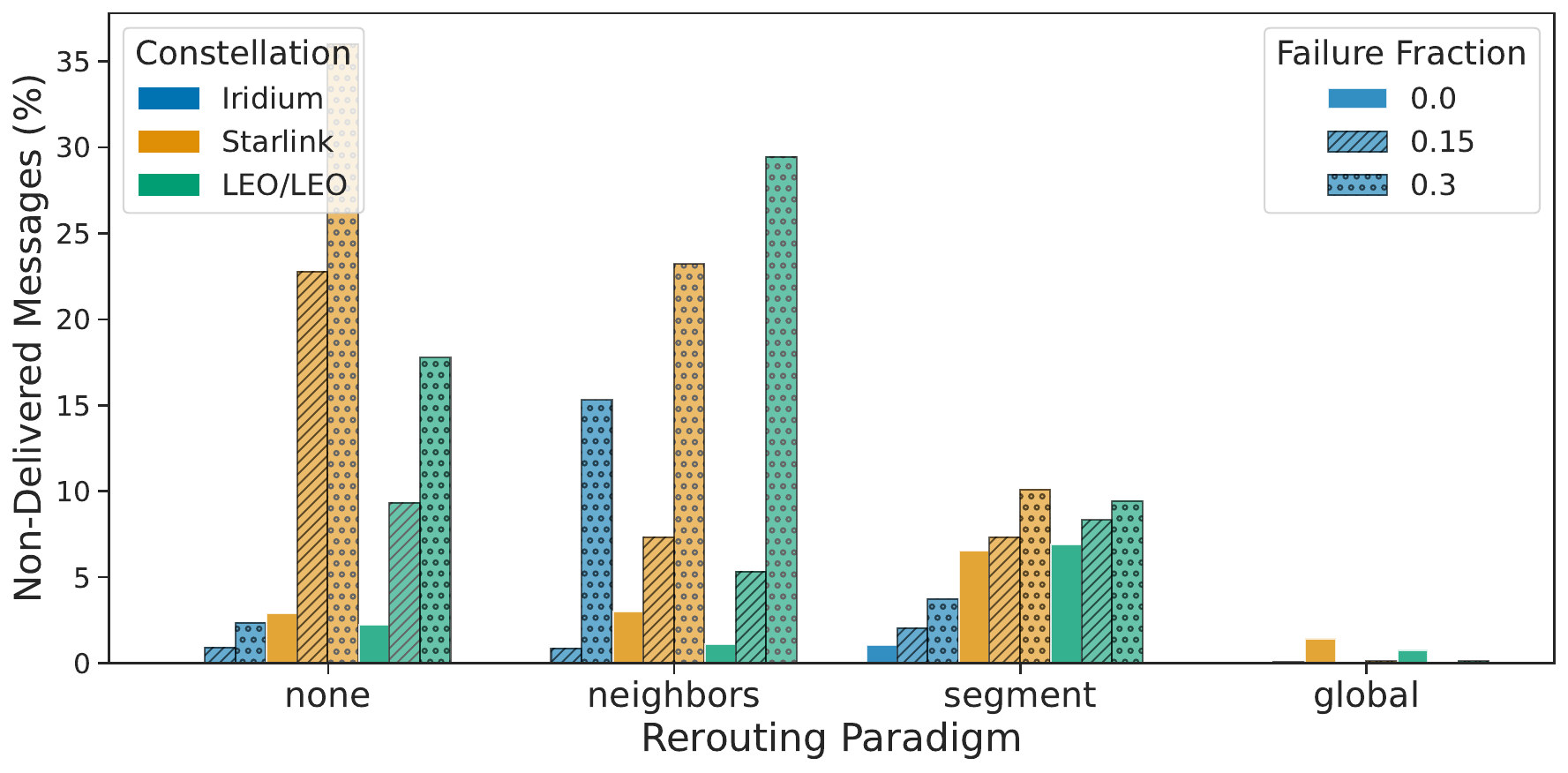}
    \caption{Percentage of non-delivered messages.\protect\footnotemark}
    \label{fig:non-delivered-messages-full}
\end{figure}
\footnotetext{For the Starlink constellation, there are messages for which no initial path exists (and are thus not sent at all), we plot statistics only for messages that were actually sent inside the satellite network}

\begin{figure}[t]
    \centering
    \includegraphics[width=0.5\textwidth]{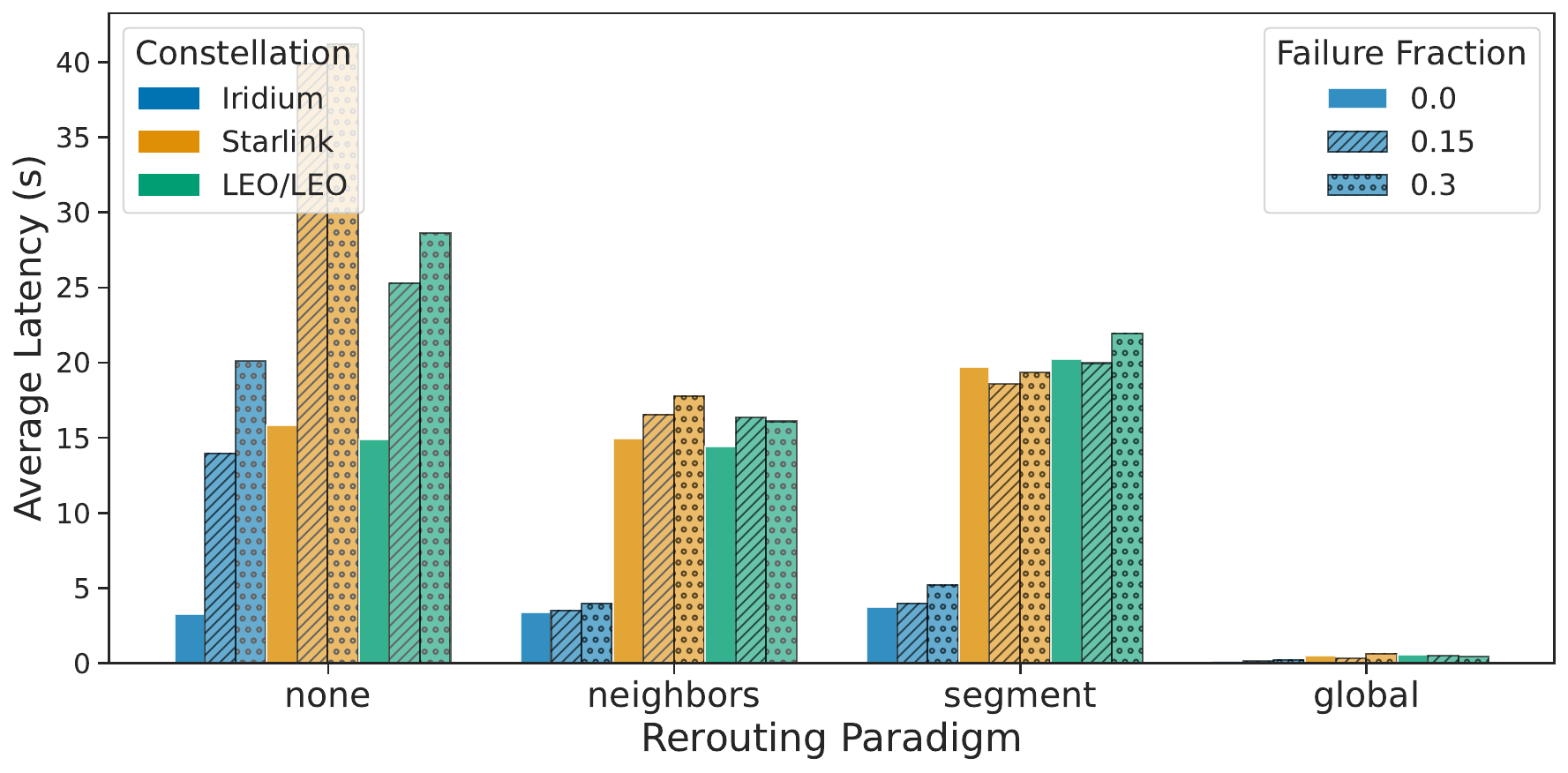}
    \caption{Average latency of the 0.97 quantile of received messages.} 
    
    \label{fig:average-latency-full}
\end{figure}

\subsubsection{Deliveries and Drop Rates}

Fig. \ref{fig:non-delivered-messages-full} demonstrates that pure source routing maintains a high delivery/low drop rate at the cost of significant latency.
Neighbor-based rerouting degrades quickly under higher failure rates, primarily due to its susceptibility to loops, causing repeated visits and eventual drops.
Segment-based rerouting generally outperforms neighbor-based in terms of delivery rates, especially under higher failure rates. However, it may still experience drops if border satellites become temporarily isolated due to surrounding link failures. 
Drops may also occur when no viable same-segment path exists (due to our constraint of using a single segment at a time), or when failure knowledge lags behind (is outdated) as a result of its delayed retrieval.

\subsubsection{Latency}
As shown in Fig.~\ref{fig:average-latency-full}, pure source routing consistently yields the highest latency, as messages must be stored when encountering failures and wait for path recovery.
Neighbor-based rerouting minimizes latency by reacting immediately to nearby failures without any signaling delay, making it the lowest-latency rerouting strategy overall.
Segment-based rerouting incurs higher delays, particularly under high failure rates. This is due to the additional signaling required to fetch up-to-date failure knowledge from border satellites, the cumulative delay growing with segment size and reroute frequency.
Global rerouting reflects idealized latency: messages that are always routed along the shortest available path given full awareness. However, this ignores the communication cost of maintaining synchronized global state.

\begin{figure}[t]
    \centering
    \includegraphics[width=0.5\textwidth]{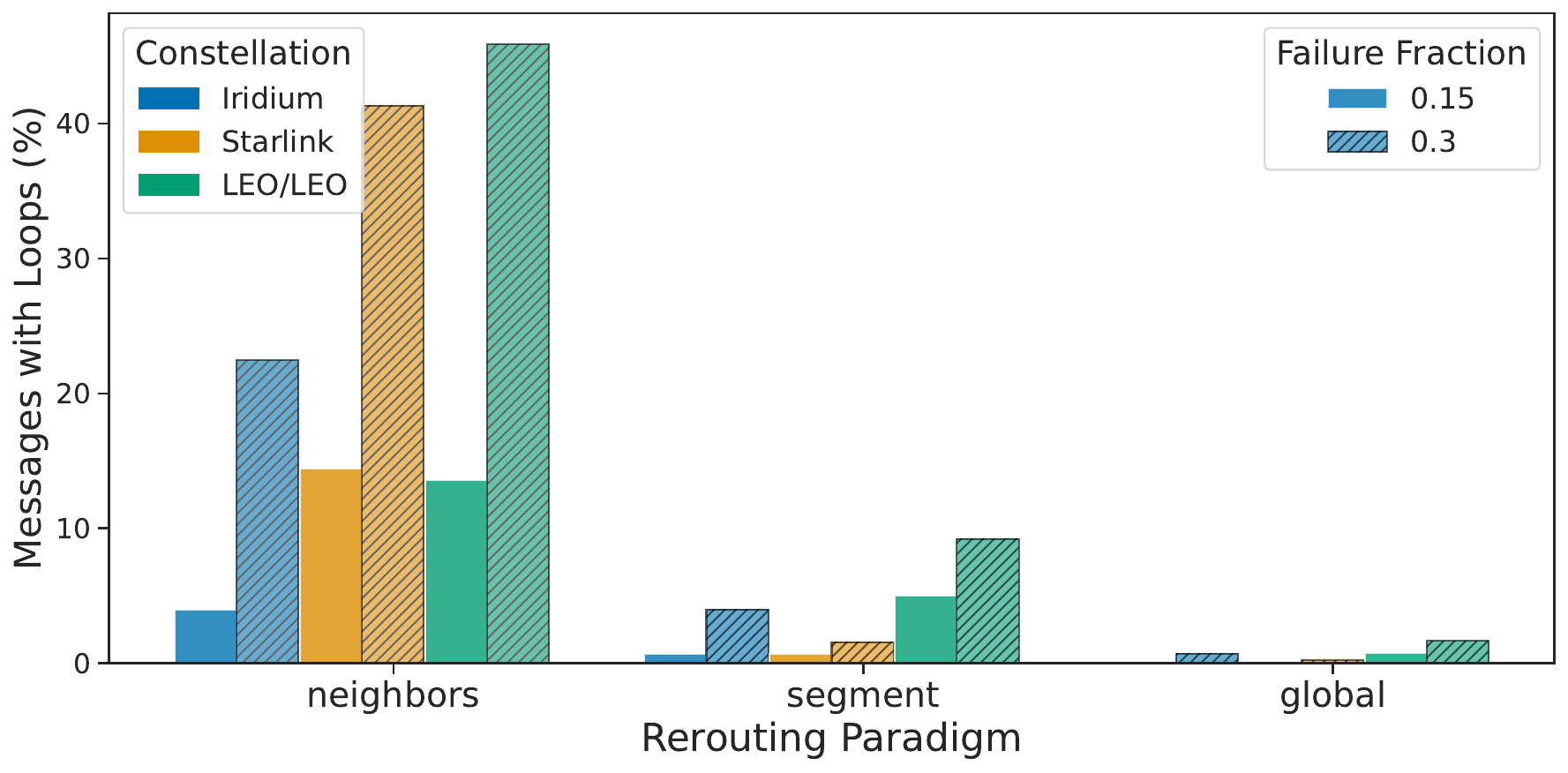}
    \caption{Messages with Loops}
    \label{fig:messages-with-loops-full}
\end{figure}

\begin{figure}[t]
    \centering
    \includegraphics[width=0.5\textwidth]{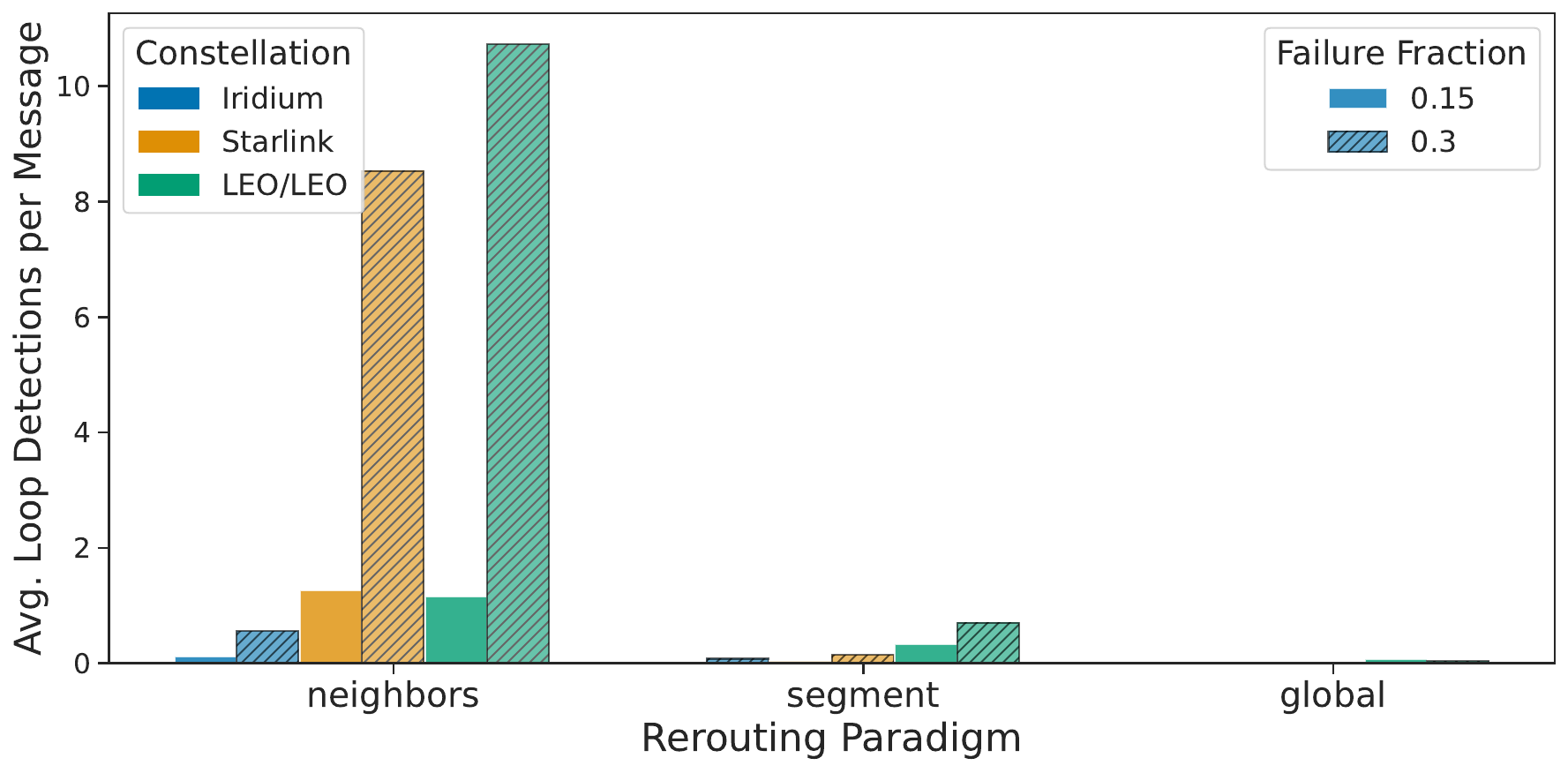}
    \caption{Average Loop Detections per message}
    \label{fig:total-loop-detections-full}
\end{figure}

\label{fig:four-metrics-combined}

\subsubsection{Routing Loops}
Fig.~\ref{fig:messages-with-loops-full} shows the percentage of messages with detected loops in all experiment runs. No loops occur in pure source routing, as rerouting is disabled. In globally informed routing, loops are rare and typically result from temporary link unavailability that forces short-term backtracking.
Neighbor-based rerouting exhibits the highest loop frequency and severity. Due to limited failure visibility and frequent rerouting, many messages revisit previously seen nodes multiple times, increasing latency and network congestion. This is illustrated also by the strong increase in the average number of loops per message, which raises more than 10-fold when increasing the fractions of failures from 0.15 to 0.3 for all three constellation settings (Fig.~\ref{fig:total-loop-detections-full}).

Segment-based rerouting, in contrast, significantly reduces both the number and depth of loops. With broader failure awareness and fewer reroute attempts, most messages avoid repeated detours.

\subsection{Targeted Failures} 

\begin{figure}[h]
    \centering
    \includegraphics[width=0.5\textwidth]{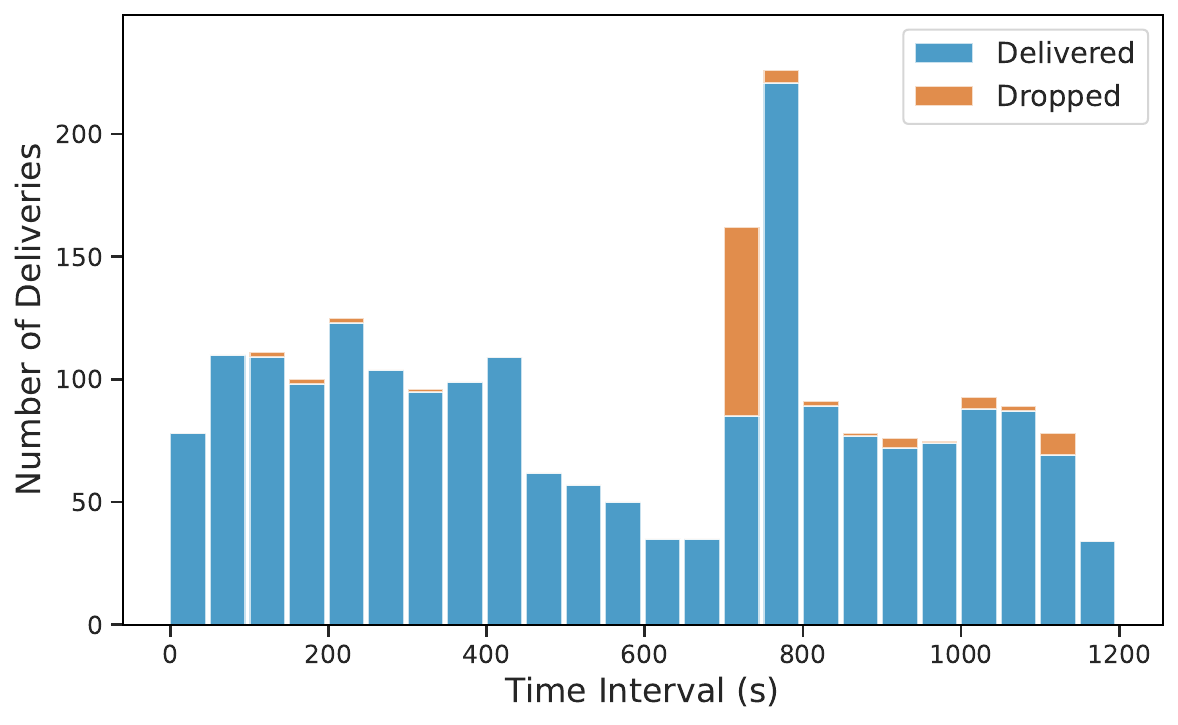}
    \caption{Effect of border satellite failures on throughput and message drops, across all constellations.}
    \label{fig:border-satellite-failures}
\end{figure}

\begin{figure}[h]
    \centering
    \includegraphics[width=0.5\textwidth]{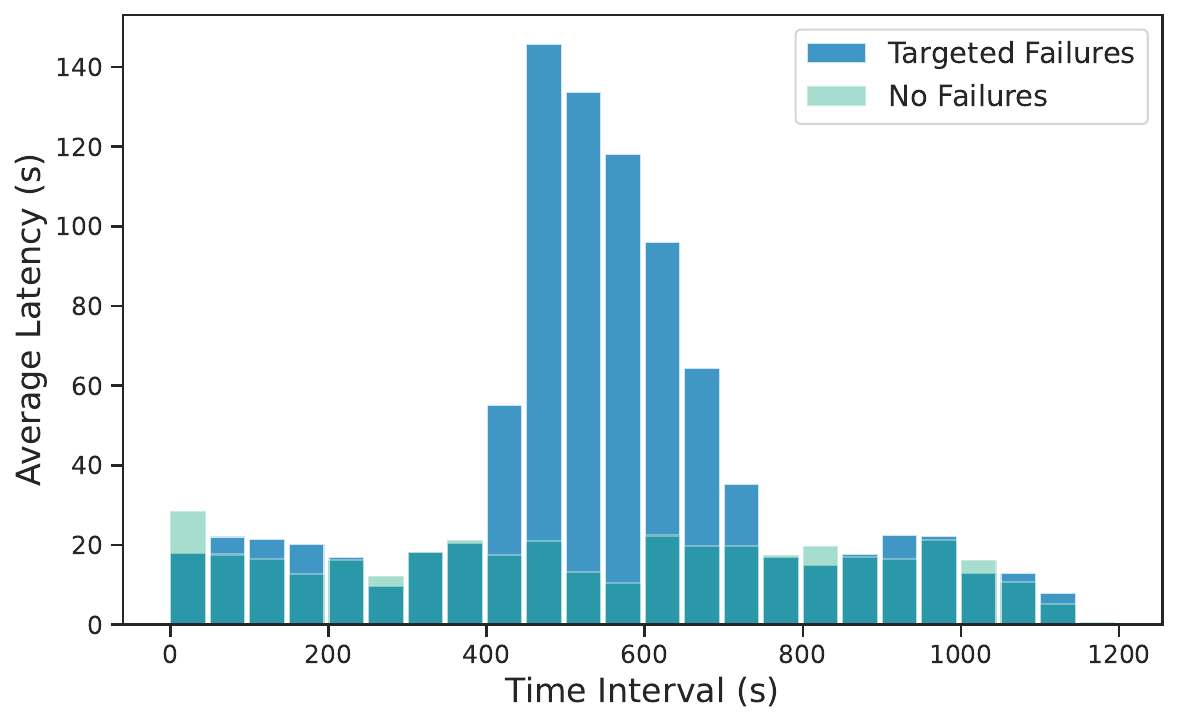}
    \caption{Effect of border satellite failures on latency, across all constellations.}
    \label{fig:border-satellite-failures-latency}
\end{figure}

\subsubsection{Deliveries and Drop Rates}
Figure \ref{fig:border-satellite-failures} shows a clear drop in deliveries once the failure of border satellites begins at 450 seconds. During this period, only messages sent between pairs of satellites within the same segment are successfully delivered. In this scenario, recovery occurs solely due to DSNS's store-and-forward functionality. This explains the sharp spike in deliveries once the routers become operational again. Note that some message drops can occur even when all border satellites are operational due to our constraint enforcing same-segment paths in each routing step.

\subsubsection{Latency}

Border satellite failures also cause an immediate and sustained spike in latency. Messages sent during the 300-second outage window are stored until the affected satellites become operational again. As a result, the delay scales with how early the message was sent during this failure period (see Fig.~\ref{fig:border-satellite-failures-latency}), highlighting the effect relative to no failures.

While DSNS’s store-and-forward mechanism enables partial recovery, the consistent spikes in latency and message drops present across all constellations highlight its limitations. These results underscore the need for more specialized mechanisms to handle border satellite failures and mitigate sustained message delays.

\subsection{Summary and Tradeoffs}

All strategies degrade under increasing failure rates, but the severity and nature of degradation depend significantly on the scope of failure awareness.

Pure source routing offers zero rerouting but exhibits poor fault tolerance in terms of performance, with messages stalled once any link on the path breaks. However, it remains viable in the Iridium constellation, if latency is of little concern.

Neighbor-based rerouting reacts quickly and with low signaling overhead, but its limited failure visibility leads to frequent loops and detours, especially under higher failure rates. Therefore, neighbor-based rerouting is not well-suited for high-traffic networks, as the increased number of reroute attempts and repeated visits to already traversed nodes would cause unmanageable congestion.

Segment-based rerouting improves delivery and loop behavior by enabling broader failure awareness within segments, though it incurs higher latency due to delayed failure information retrieval---on average, 54\% of a message's end-to-end latency stems from signaling overhead. Segment-based rerouting is particularly well-suited to Walker-style constellations, as it requires only a single partitioning of the network—leveraging the relative stability of satellite neighborhoods. Furthermore, it consistently performs better than neighbor-based rerouting, which makes it generally preferable for use in Walker-style constellations.

Global rerouting achieves optimal performance with full awareness but remains impractical at scale due to its synchronization demands.

Each routing paradigm offers tradeoffs between scalability, responsiveness, and robustness. Our results show a gradual, measurable improvement in delivery rates, latency and message loops as failure-awareness increases, underscoring the value of scoped, responsive awareness in large-scale satellite networks.

\section{Current Limitations and Future Work}

Our work lays the foundation for simulating and evaluating scoped failure-aware routing in large-scale satellite networks. However, several important directions remain open for future exploration.

\subsection{Resilience Mechanisms and Loop Handling}
Currently, packets are allowed to visit a given satellite more than once, and only infinite loops are prevented. Future work could improve this by implementing more flexible loop-handling strategies, combining more general loop prevention with caching of fallback paths, or triggering controlled retransmissions. Dynamic failure recovery protocols such as spanning tree variants or hop-by-hop signaling, as showcased in \cite{feng2020elastic} may also reduce unnecessary loss and improve convergence.

\subsection{Localized Routing Enhancements}
Neighbor-based rerouting could benefit from lightweight path caching or failover mechanisms, especially since rerouting scope is limited to nearby nodes. Because rerouting typically occurs within a 3-4 hop radius, storing localized fallback paths may significantly reduce delay without increasing global state. Similarly, segment-based rerouting could be extended with multipath support and fallback border satellites, improving fault tolerance without requiring full-network visibility.

\subsection{Dynamic Segmentation and Border Satellite Redundancy.}
While our current segment partitioning is static, dynamic segmentation could be explored to adapt to longer-term topology changes, or topologies different from classical LEO networks. This would involve periodically reclustering nodes and updating border satellite assignments, potentially with techniques such as Discrete Particle Swarm Optimization (DPSO) \cite{dai2021intelligent}. Future versions could also introduce redundant central or distributed routers within each segment, improving resilience under targeted attacks and reducing signaling overhead and latency.

\subsection{Cross-Domain Integration.}
Routing performance and PKI management are tightly coupled in space networks. Exploring the effects of routing failure on certificate freshness, revocation propagation, or quorum protocols would extend our model toward full-system resilience. Moreover, simulating advanced failure types (e.g., jamming, false-state injection, or Byzantine nodes) would bring the system closer to adversarially robust operation.

\subsection{Scalability and Realistic Topologies.}
Our evaluation framework could be extended to larger, multi-layer constellations (e.g., LEO-MEO-LEO), where inter-layer links introduce additional complexity. In such cases, techniques like distributed control planes, localized knowledge caches, or hybrid routing policies (e.g., global source routing with local segment failover) may be required. 
Ultimately, our simulator serves as a flexible foundation for future research on resilient, scalable, and secure routing in satellite networks due to improved loop resolution, smarter failure awareness mechanisms, and more dynamic reconfiguration of the control and data planes.

\section{Conclusion}

In this work, we investigated how the scope of failure-awareness influences routing resilience in LEO satellite networks, particularly under adversarial and fault-prone conditions. By implementing and evaluating four scoped rerouting strategies in the Deep Space Network Simulator, we showed that broader awareness enables more reliable message delivery and stronger loop avoidance, even amid targeted disruptions. Segment-based rerouting may provide a robust, scalable compromise, offering improved security posture without the prohibitive overhead of global state. These results highlight the importance of designing satellite routing protocols with security-resilient primitives that respond adaptively to both random failures and intentional attacks.

\printbibliography

\end{document}